\documentclass[epj,twocolumn]{webofc}
\usepackage[varg]{txfonts}   
\woctitle{LHCP 2013}
\begin{document}
\title{Direct Photon Results from CDF}
%
%

\author{Tingjun Yang \inst{1}\fnsep\thanks{\email{tjyang@fnal.gov}} for the CDF Collaboration}

\institute{Fermilab, Batavia, IL, USA
          }

\abstract{%
 Direct (prompt) photon  production is a field of very high interest in hadron colliders. It provides probes to search for new phenomena and to test QCD predictions. In this article, two recent cross-section results for direct photon production using the full CDF Run II data set are presented: diphoton production and photon production in association with a heavy quark.
}
\maketitle
\section{Introduction}
\label{intro}
Direct photon ($\gamma$) production in hadron colliders provides a clean probe to test quantum chromodynamics (QCD) predictions. A better understanding of such processes can improve the background modeling in the searches for new physics with photon final state. In this article, two recent cross-section results for direct photon production in $p\bar{p}$ collisions at $\sqrt{s} = 1.96$ TeV are presented: diphoton production \cite{Aaltonen:2012jd} and photon production in association with a heavy quark $Q$ ($b$ or $c$) \cite{Aaltonen:2013coa}. Both results use the full data set collected by the CDF II detector. 

The CDF II detector \cite{Acosta:2004hw} has a cylindrical geometry with approximate forward-backward and azimuthal symmetry. It contains a tracking system consisting of silicon microstrip detectors and a cylindrical open-cell drift chamber immersed in a 1.4 T magnetic field parallel to the beam axis. The silicon subsystem is used for reconstructing charged-particle trajectories (tracks) and heavy-flavor-decay vertices displaced from the primary interaction point. Electromagnetic (EM) and hadronic calorimeters surrounding the tracking system with pointing-tower geometry are used to measure photon energies. At a depth approximately corresponding to the maximum development of the EM shower, the EM calorimeters contain fine-grained detectors (central electromagnetic strip chambers) that measure the shower profile. Drift chambers and scintillators located outside the calorimeters identify muons. 

\section{Cross Section for Direct Diphoton Production}
\label{sec-1}
Precise measurements of the production cross sections for diphotons are important for the searches for new phenomena, such as extra spatial dimensions, and for improvements in the precision of the measurements of the production cross section and the decay branching ratio of the Higgs boson into a photon pair. Diphoton production is also used to test QCD both in the perturbative scheme (pQCD) and in nonperturbative schemes, such as soft-gluon resummation methods and photons from quark fragmentation. This article presents the final diphoton measurements from CDF using the full data set collected in 2001-2011 corresponding to a total integrated luminosity of 9.5 fb$^{-1}$. 

Inclusive diphoton events are selected online by requiring two isolated electromagnetic clusters with $E_{T} > 12$ GeV each or two electromagnetic clusters with $E_{T} > 18$ GeV and no isolation requirement. In the offline analysis additional requirements are imposed to identify a sample rich in prompt photons. The pseudorapidity of each photon in the event is restricted to the region $|\eta| < 1$. The photon transverse energy is required to exceed 17 GeV for the first photon and 15 GeV for the second photon. The transverse energy measured by the calorimeter in an isolation cone with a radius in $\eta-\phi$ space of 0.4 around each photon is required not to exceed 2 GeV.

The background from events where one or both reconstructed photons are misidentified jets is subtracted with a $4\times4$ matrix technique using the track isolation as the discriminant between the signal and background \cite{Aaltonen:2011vk}. The differential cross section for diphoton production is obtained from the histogram of the estimated signal yield as a function of each relevant kinematic variable. The average cross section in a bin is determined by dividing the yield by the product of the trigger efficiency, the selection efficiency and acceptance, the integrated luminosity, and the bin size.

The experimental results are compared with six theoretical calculations: 
(i) the fixed NLO predictions of the {\sc diphox} program, including nonperturbative parton fragmentation into photons at NLO, (ii) the predictions of the {\sc resbos} program where the cross section is accurate to NLO, but also has an analytical initial-state soft-gluon resummation, (iii) the predictions of the {\sc pythia} parton-shower program including photons radiated from initial- and final-state quarks, (iv) the fixed NLO predictions of the {\sc mcfm} program, including nonperturbative parton fragmentation into photons at LO, (v) the fixed next-to-next-to-leading order (NNLO) predictions of a recent calculation, and (vi) the predictions of the {\sc sherpa} program, based on a matrix element calculation merged with the parton shower model.

The measured cross section for diphoton production integrated over the
acceptance is $12.3\pm 0.2_{\rm stat}\pm 3.5_{\rm syst}$$~$pb. The predictions
for the integrated cross section are $10.6\pm 0.6$$~$pb from {\sc diphox}, $11.3\pm 2.4$$~$pb from {\sc resbos}, $9.2$$~$pb from {\sc pythia}, $12.4\pm 4.4$$~$pb from {\sc sherpa},
$11.5\pm 0.3$$~$pb from {\sc mcfm}, and $11.8^{+1.7}_{-0.6}$$~$pb from the NNLO
calculation. The {\sc sherpa} scale uncertainty is large because it
also accounts for parton shower. The {\sc pythia} uncertainty is unreported. All predictions are consistent with the measurement.

\begin{figure*}
\centering
\includegraphics[width=7.5cm]{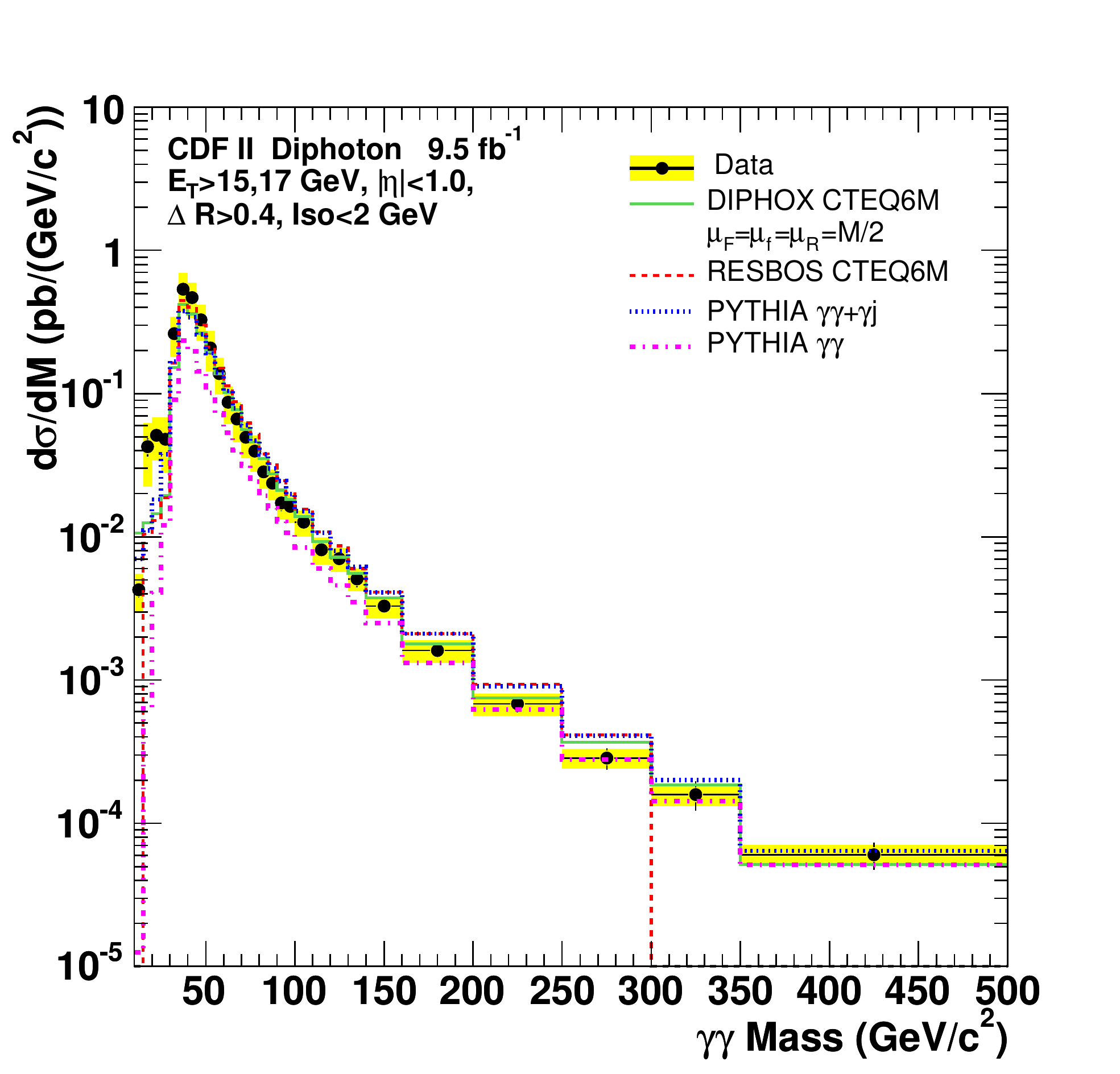}\hspace*{0cm}
\includegraphics[width=7.5cm]{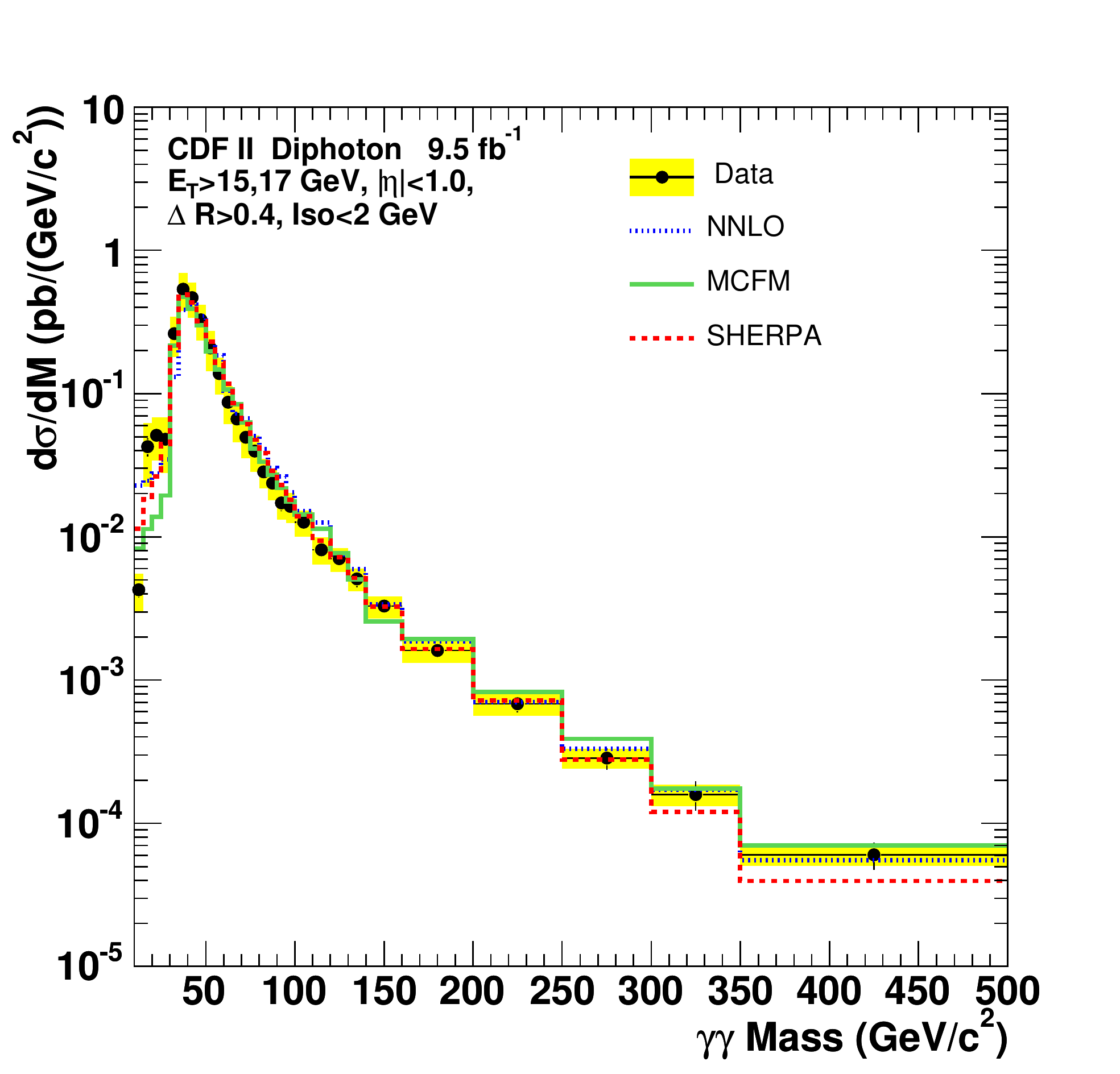}
\vspace*{0cm}
\includegraphics[width=7.5cm]{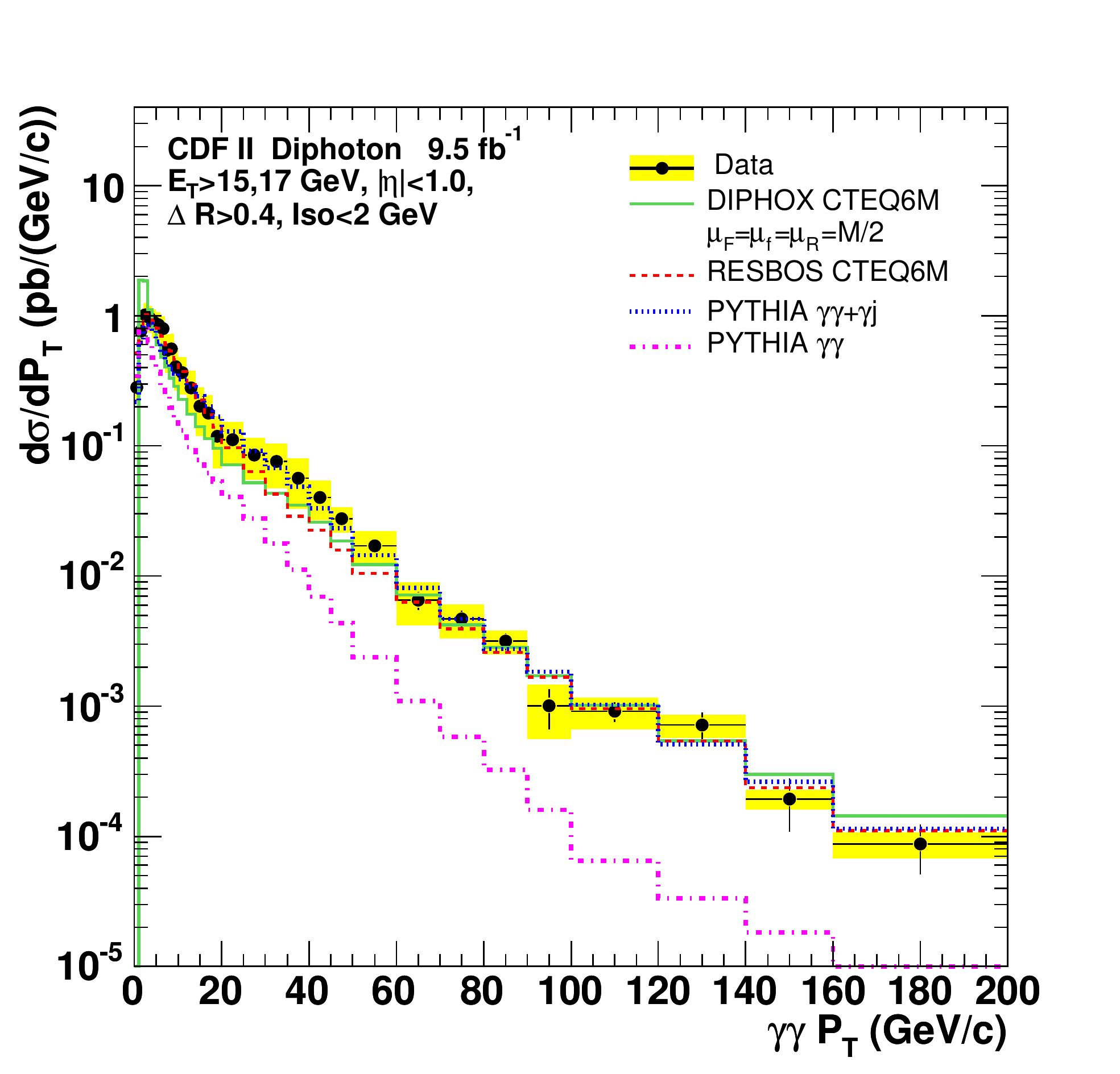}\hspace*{0cm}
\includegraphics[width=7.5cm]{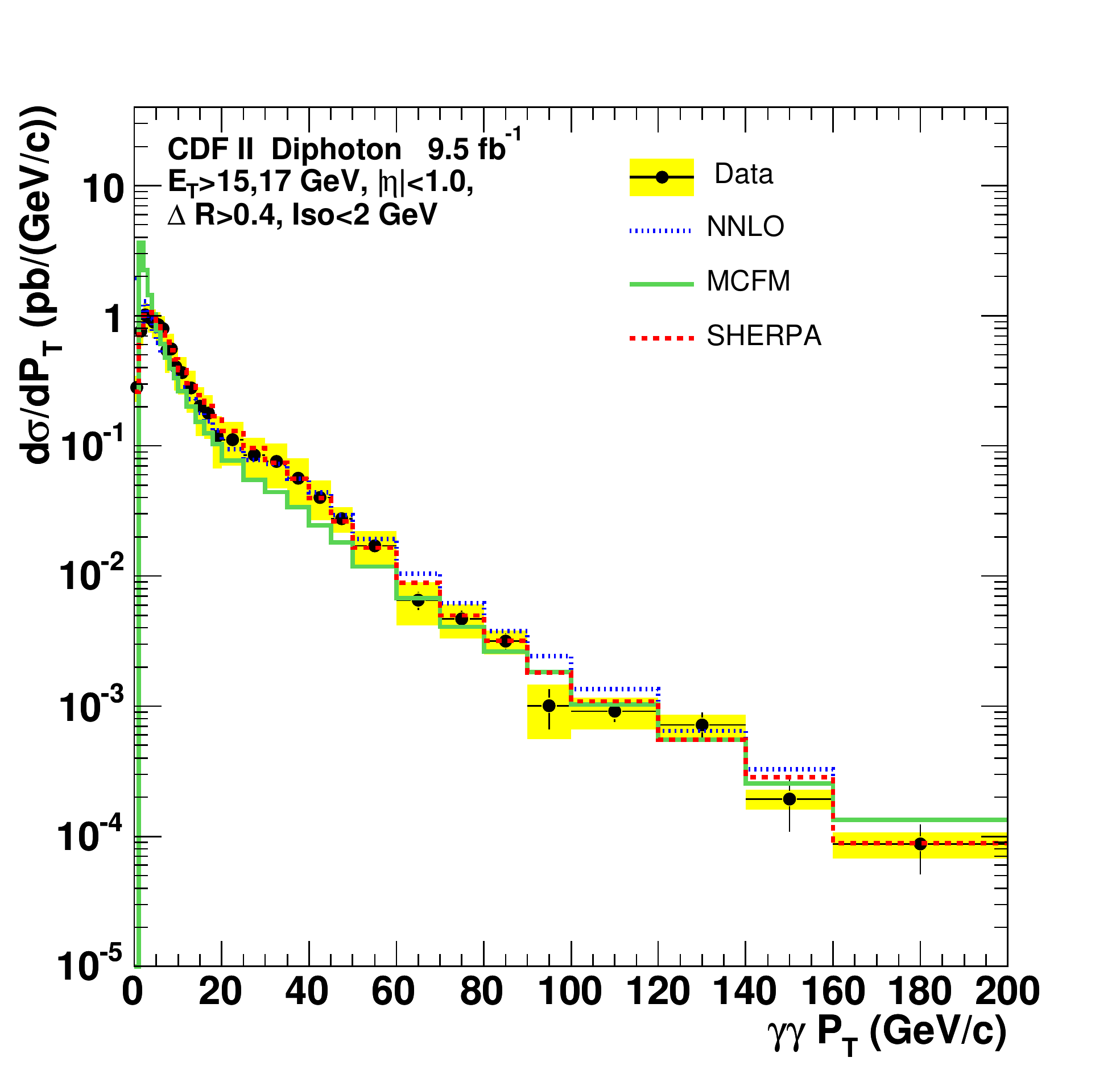}
\vspace*{0cm}
\includegraphics[width=7.5cm]{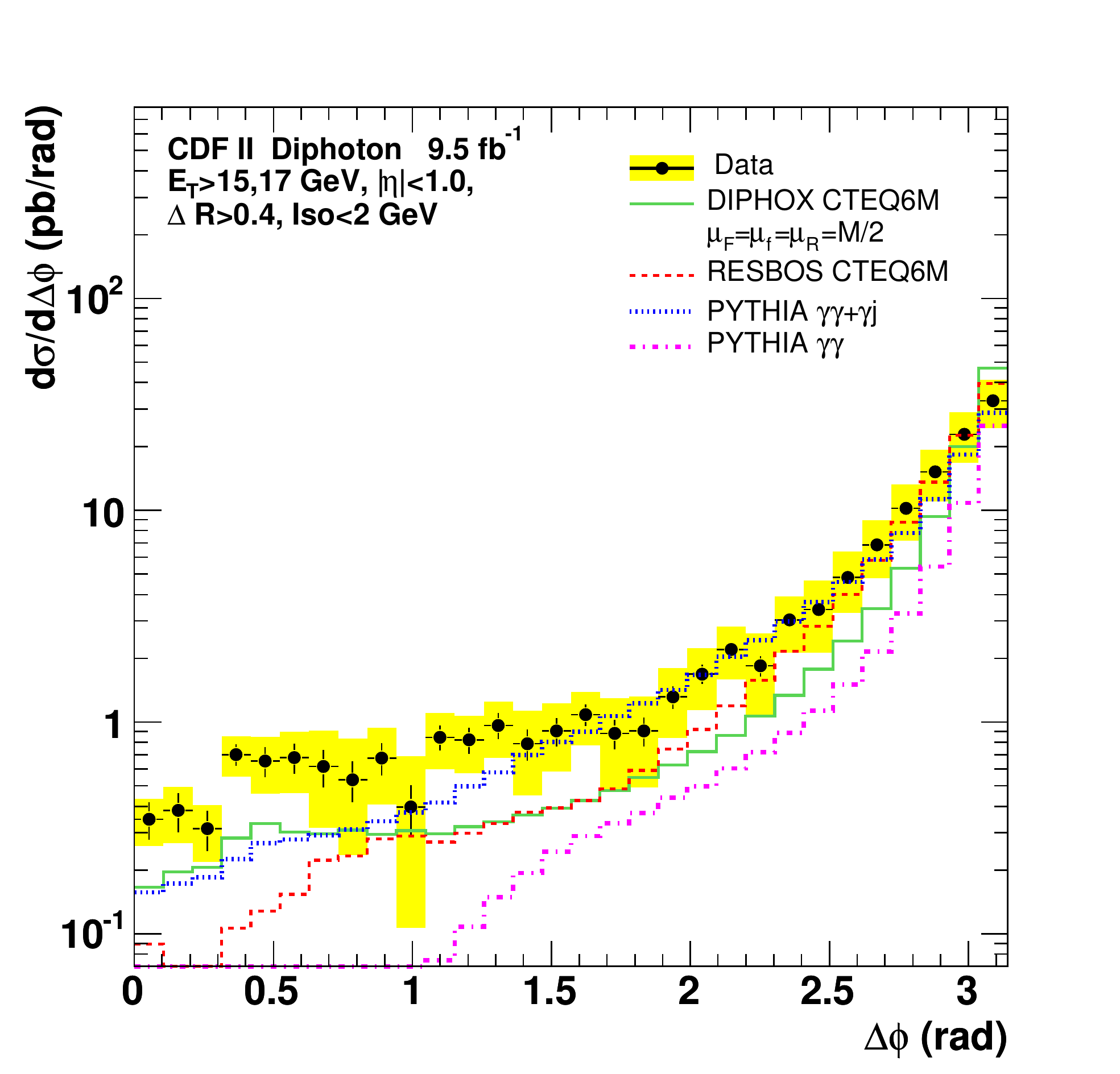}\hspace*{0cm}
\includegraphics[width=7.5cm]{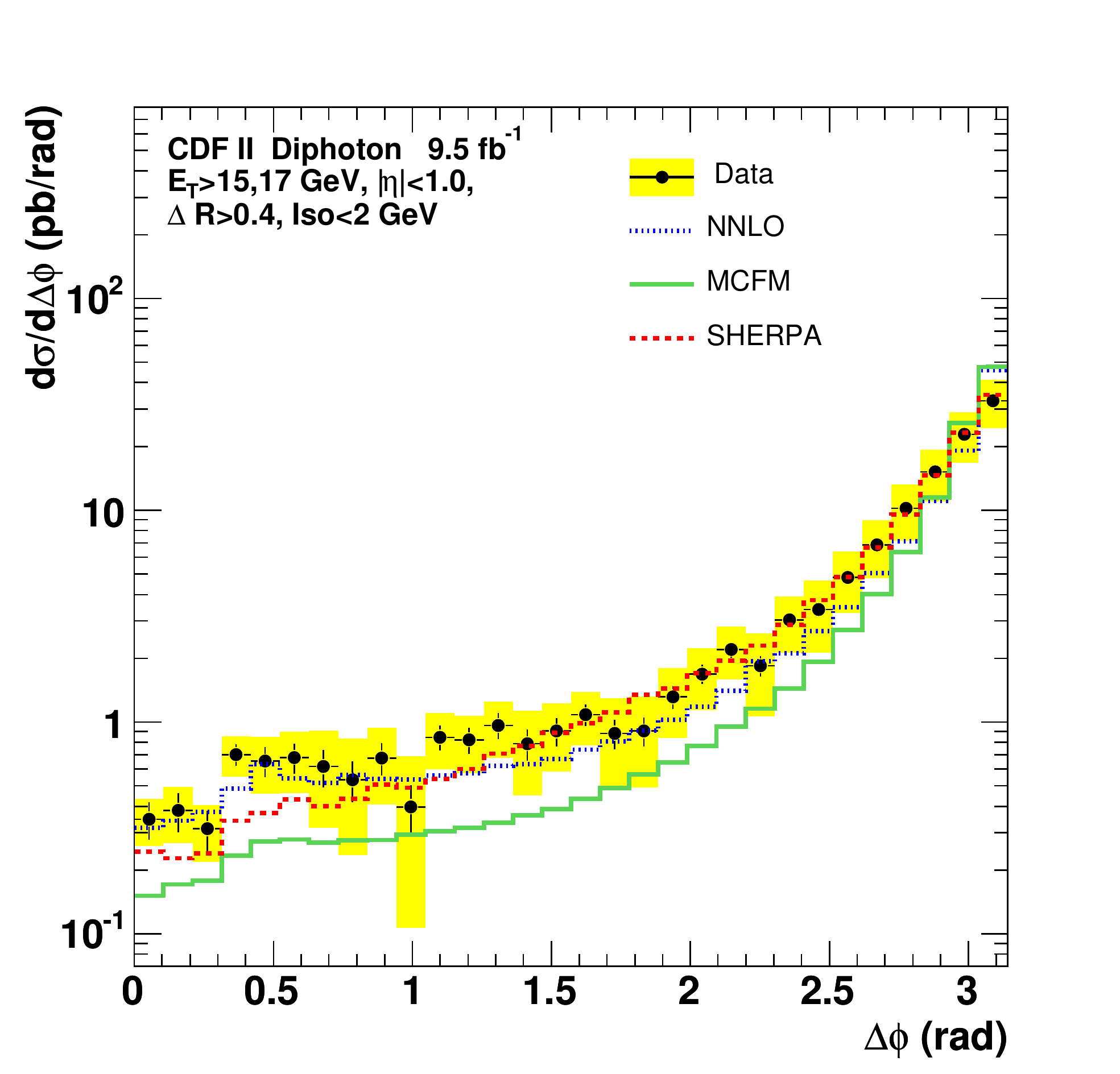}
\vspace*{-0.1cm}
\caption{Measured differential cross sections as functions of the diphoton
mass (top) and transverse momentum (middle), and of the azimuthal difference
between the photon directions (bottom), compared with six theoretical
predictions discussed in the text. The shaded area around the data points indicates the total systematic uncertainty of the measurement.\label{fig-1}}
\end{figure*}

Figure~\ref{fig-1} shows the comparisons between the observed and predicted distributions in diphoton mass $M$, transverse momentum $P_{T}$ of the photon pair, and azimuthal separation $\Delta\phi$ between the momenta of the two photons in the event. All predictions for the mass distribution show a reasonable agreement with the data for all calculations above the maximum at 30 GeV/$c^{2}$ except the {\sc pythia} $\gamma\gamma$ calculation. All predictions underestimate the data rate around and below the maximum. In the $P_{T}$ spectrum, the {\sc pythia}, {\sc diphox}, {\sc resbos} and {\sc mcfm} predictions underestimate the data in the region between 30 and 60 GeV/$c$, where the contribution from quark fragmentation is important. Both NNLO and {\sc sherpa} predictions describe the data fairly well in this region. For $P_{T} < 20$ GeV/$c$, where soft-gluon radiation becomes important, the {\sc resbos}, {\sc pythia} and {\sc sherpa} predictions provide a good description of the data because of the resummation of multiple soft-gluon emission amplitudes through either analytical calculation or parton showering. The fixed-order predictions diverge in the limit of vanishing $P_{T}$. The {\sc resbos}, {\sc pythia} and {\sc sherpa} predictions show a good agreement with the data at larger $\Delta\phi$, where the diphoton system acquires substantial transverse momentum due to multiple soft-gluon emission. The NNLO calculation is the only prediction consistent with the data in the low $\Delta\phi$ tail, which contains photon pairs with very low mass and relatively high $P_{T}$. 

\section{Cross Section for Direct Photon Production in Association with a Heavy Quark}
\label{sec-2}
The cross section for direct photon production in association with a heavy quark in hadronic collisions provides valuable information on the probability distributions of partons inside the initial-state hadrons. At photon transverse energy $E_{T}^{\gamma}$ smaller than 100 GeV, such events are produced predominantly by the Compton scattering process, while at higher energies the dominant process is quark-antiquark annihilation with gluon ($g$) splitting to heavy quarks. It is conventional to assume that the charm ($c$) and bottom ($b$) quarks in the proton arise only from gluon splitting. However, there are other models that allow the existence of intrinsic heavy quarks in the proton. A cross-section measurement of $\gamma+Q+X$ ($X$ can be any final-state particle) production provides information on the heavy-quark and gluon parton distribution functions (PDFs) and on the rate of final-state gluon splitting to heavy quarks. This article presents the final CDF measurements of the cross sections for photon with heavy-flavor jets, using the full data set from 9.1 fb$^{-1}$ of integrated luminosity, exploring $E_{T}^{\gamma}$ up to 300 GeV.

The photo-plus-heavy-jet events are selected online by requiring at least one energy cluster consistent with a photon in the final state. The offline event selection required each event to have at least one photon candidate that has pseudorapidity in the fiducial region of the central calorimeter ($|\eta| < 1$). Photon candidates are required to have $E_{T}^{\gamma} > 30$ GeV. An artifical neural network (ANN) is constructed to reduce background. At least one jet must be present in each event. Jets are reconstructed using the JETCLU algorithm with a cone radius 0.4. We select jets that have $E_{T} > 20$ GeV and $|\eta| < 1.5$. At least one jet is required to be classified as a heavy-flavor jet using a secondary-vertex tagger. The selected jet is required to be reconstructed in a volume outside a cone with a radius in $\eta-\phi$ space of 0.4 surrounding the photon candidate. 

\begin{figure}
\centering
\includegraphics[width=7.5cm,clip]{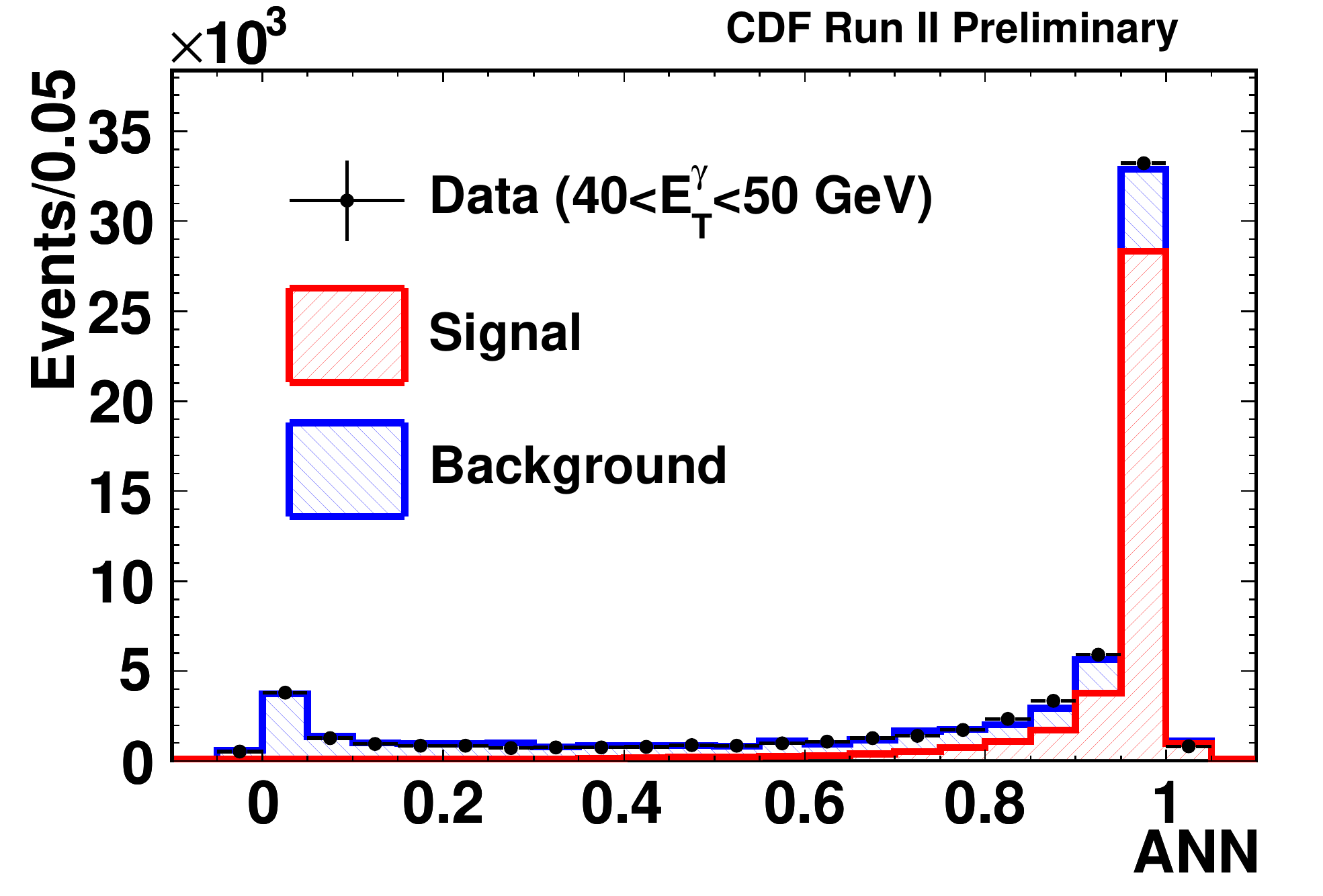}
\caption{The fit to the ANN distribution for photon candidates with $E_{T}^{\gamma}$ between 40 and 50 GeV. The points are data, and the stacked, shaded histograms represent the estimated contributions from the fit of the prompt photon signal and false photon background.}
\label{fig-2}       
\end{figure}

There are two main background sources: jets misidentified as photons (false photons) and light-flavor jets mimicking heavy-flavor jets. To estimate the rate of false photons, the photon ANN distribution in data is fitted to a linear combination of templates for photons and jets, obtained from simulated samples. A fit is performed in each $E_{T}^{\gamma}$ interval to get the prompt photon fractions (purities). One example fit is shown in Figure~\ref{fig-2}.

\begin{figure}
\centering
\includegraphics[width=7.5cm,clip]{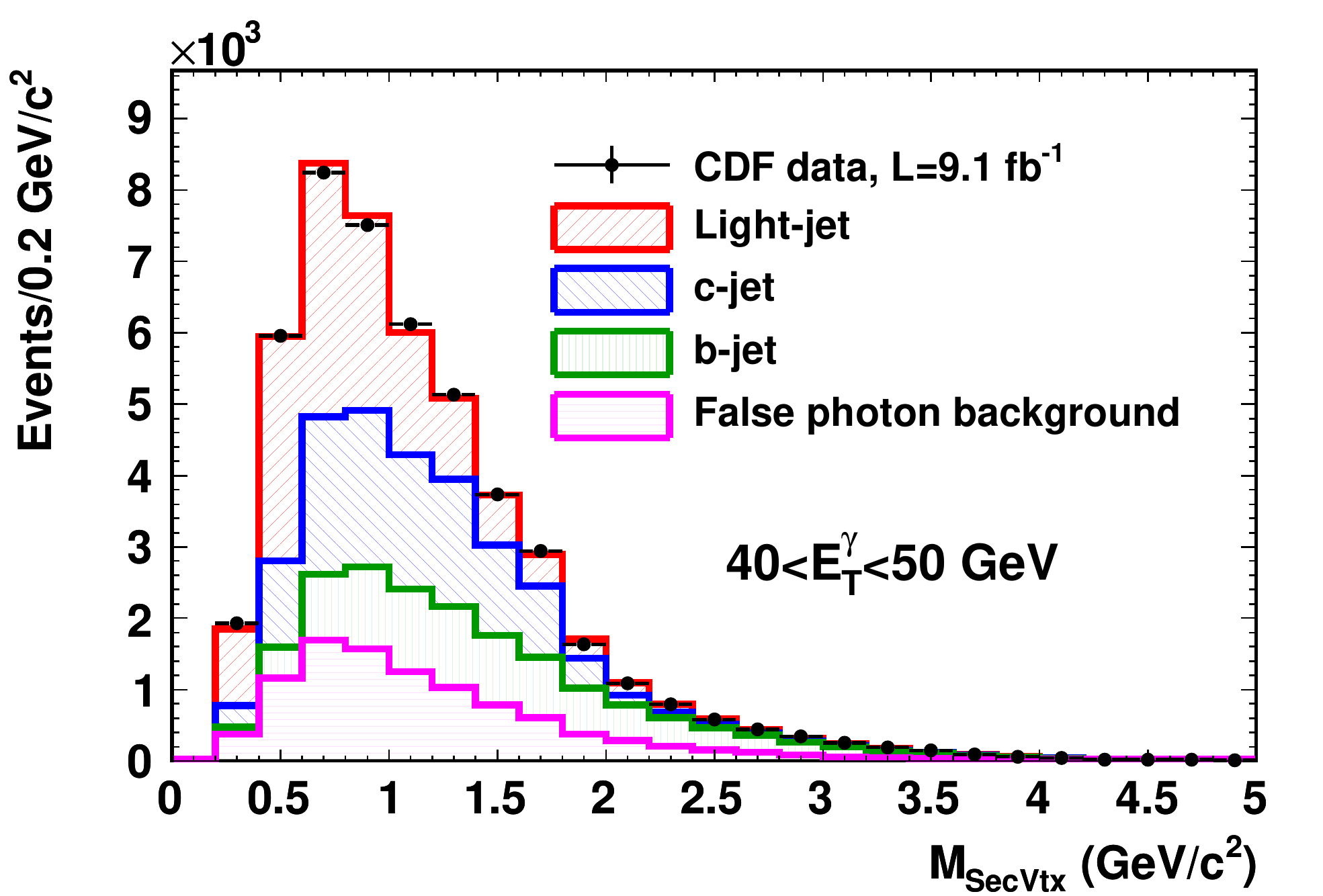}
\caption{Distribution of the secondary-vertex mass of tagged jets for photon candidates with $40 < E_{T}^{\gamma} < 50$ GeV. The points are data, and the stacked, shaded histograms represent the estimated contributions from the fit of the $b$-, $c$-, and light-quark jets and false photon background.}
\label{fig-3}       
\end{figure}

Backgrounds to heavy-flavor jets arise from light-flavor jets where random combinations of tracks mimic a displaced vertex. The fractions of $b$- and $c$-jets are determined by fitting the invariant mass ($M_{\rm SecVtx}$) of the system of charged particles, assumed to be pions, originating at the secondary vertex, using the templates for $b$-, $c$-, and light-quark jets constructed with a simulated sample. The contribution to the $M_{\rm SecVtx}$ distribution from events with a false photon is modeled using dijet data. Figure~\ref{fig-3} shows the result of the fit for $E_{T}^{\gamma}$ between 40 and 50 GeV, as an example.

The differential cross section is obtained from the histogram of the estimated signal yield as a function of photon $E_{T}$. The average cross section in a bin is determined by dividing the yield by the product of the trigger efficiency, the selection efficiency and acceptance, the integrated luminosity, and the bin size. The experimental results are compared with four theoretical calculations: (i) the fixed NLO calculations, including direct photon production subprocesses and subprocesses where the photon is emitted from parton fragmentation, both at ${\cal O}(\alpha\alpha_{s}^2)$, (ii) the $k_{T}$-factorization calculations, including ${\cal O}(\alpha\alpha_{s}^2)$ off-shell amplitudes of gluon-gluon fusion and quark-(anti)quark interaction subprocesses, and the $k_T$-dependent (i.e., unintegrated) parton distributions, where $k_T$ denotes the transverse momentum of the parton, (iii) the {\sc sherpa} calculations, including all the tree-level matrix-element diagrams  with one photon and up to three jets, with at least one $b$ jet or $c$ jet in the explored kinematic region, (iv) the {\sc pythia} calculations, including the $2\rightarrow2$ matrix-element subprocesses $gb\rightarrow\gamma b$ and $q\bar{q}\rightarrow\gamma g$ with $g\rightarrow b\bar{b}$ and $g\rightarrow c\bar{c}$ splittings in the parton shower. Previous studies \cite{Abbott:1999se} showed that the contribution of gluon splitting to heavy flavor has to be approximately doubled over expectations from the leading-order {\sc pythia} generator to reproduce the data. Hence, we also show predictions that include a double gluon-splitting rate to heavy flavors.

\begin{figure*}
\centering
\includegraphics[width=7.5cm]{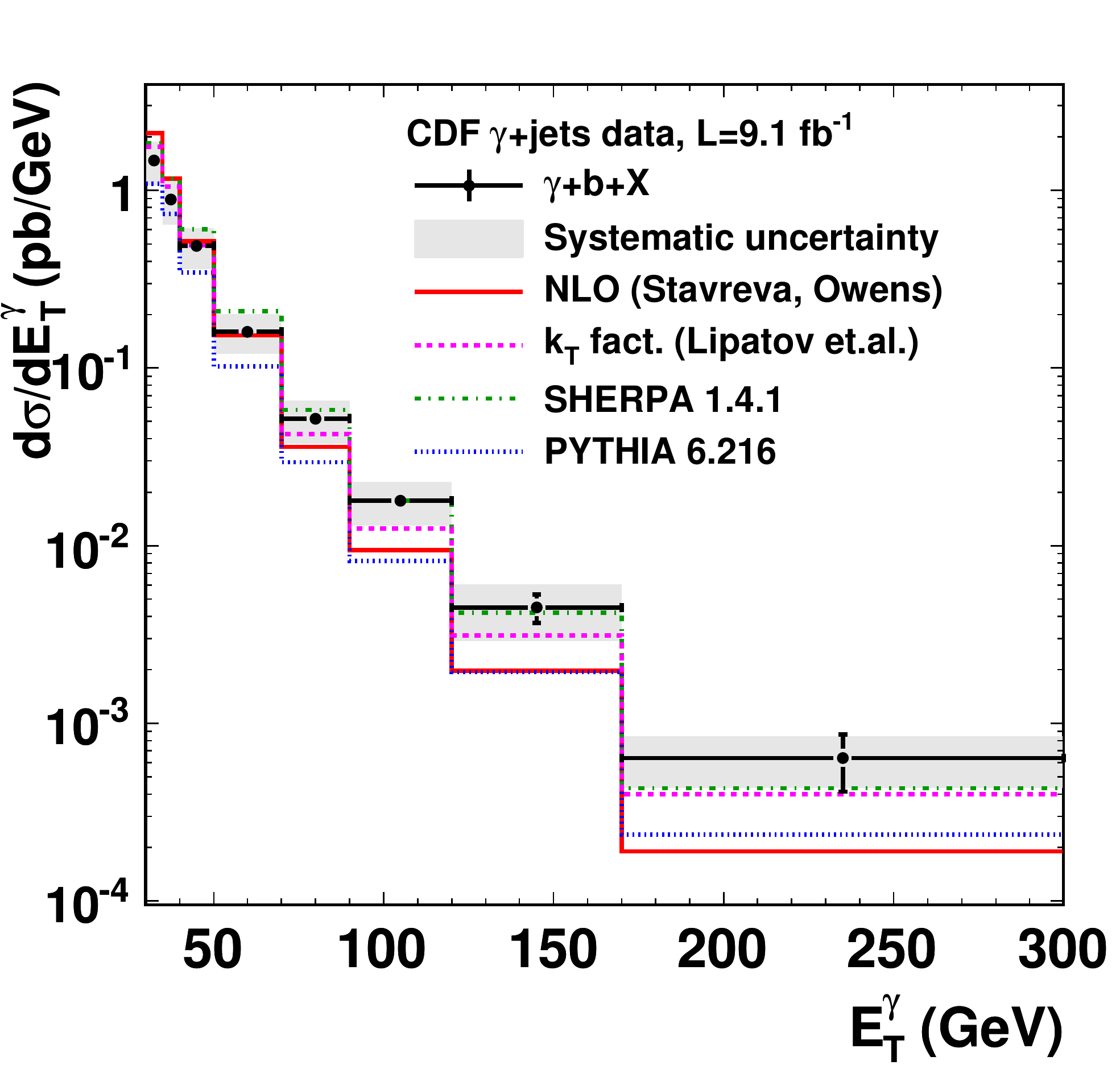}\hspace*{0cm}
\includegraphics[width=7.5cm]{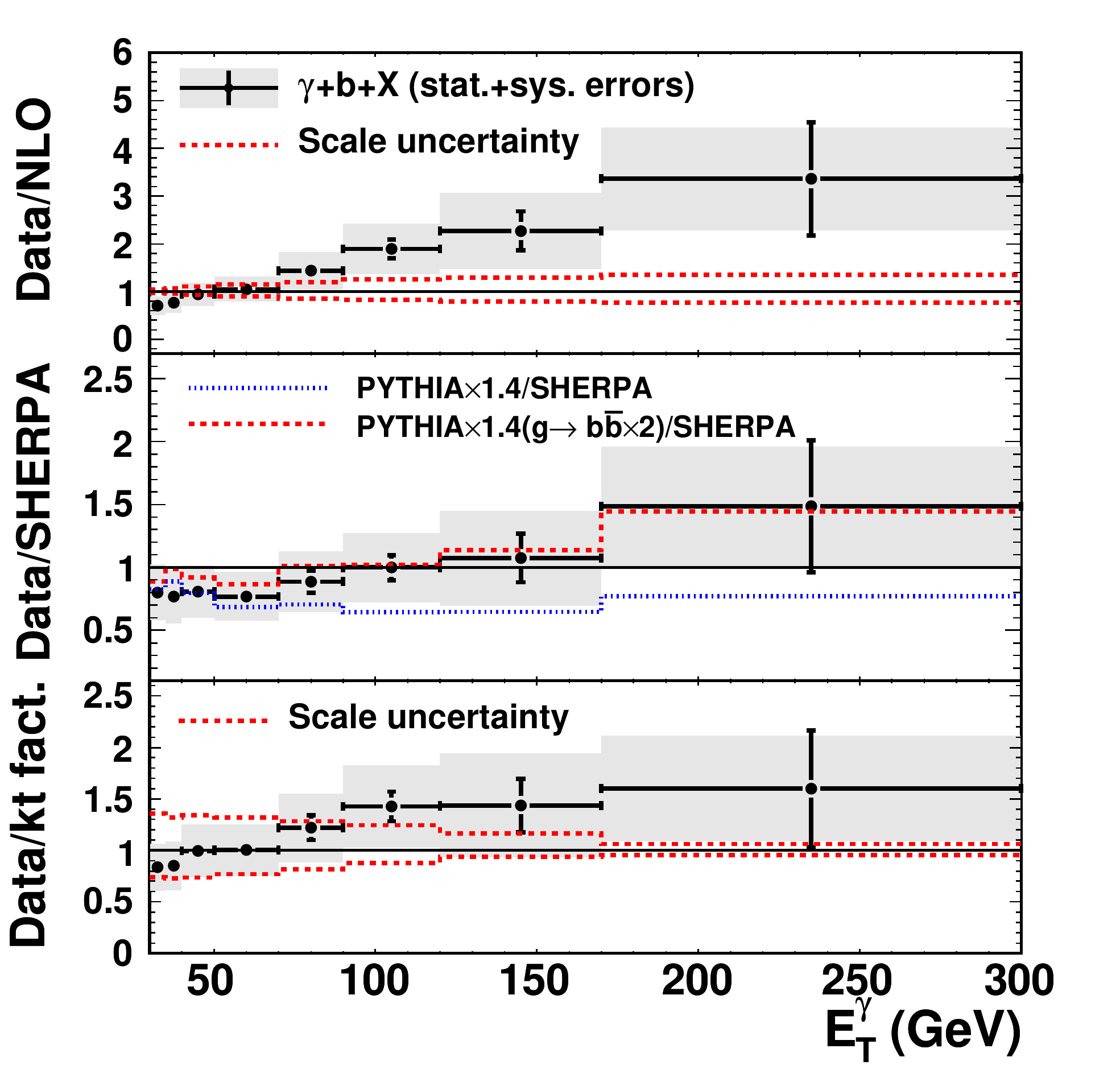}
\vspace*{0cm}
\includegraphics[width=7.5cm]{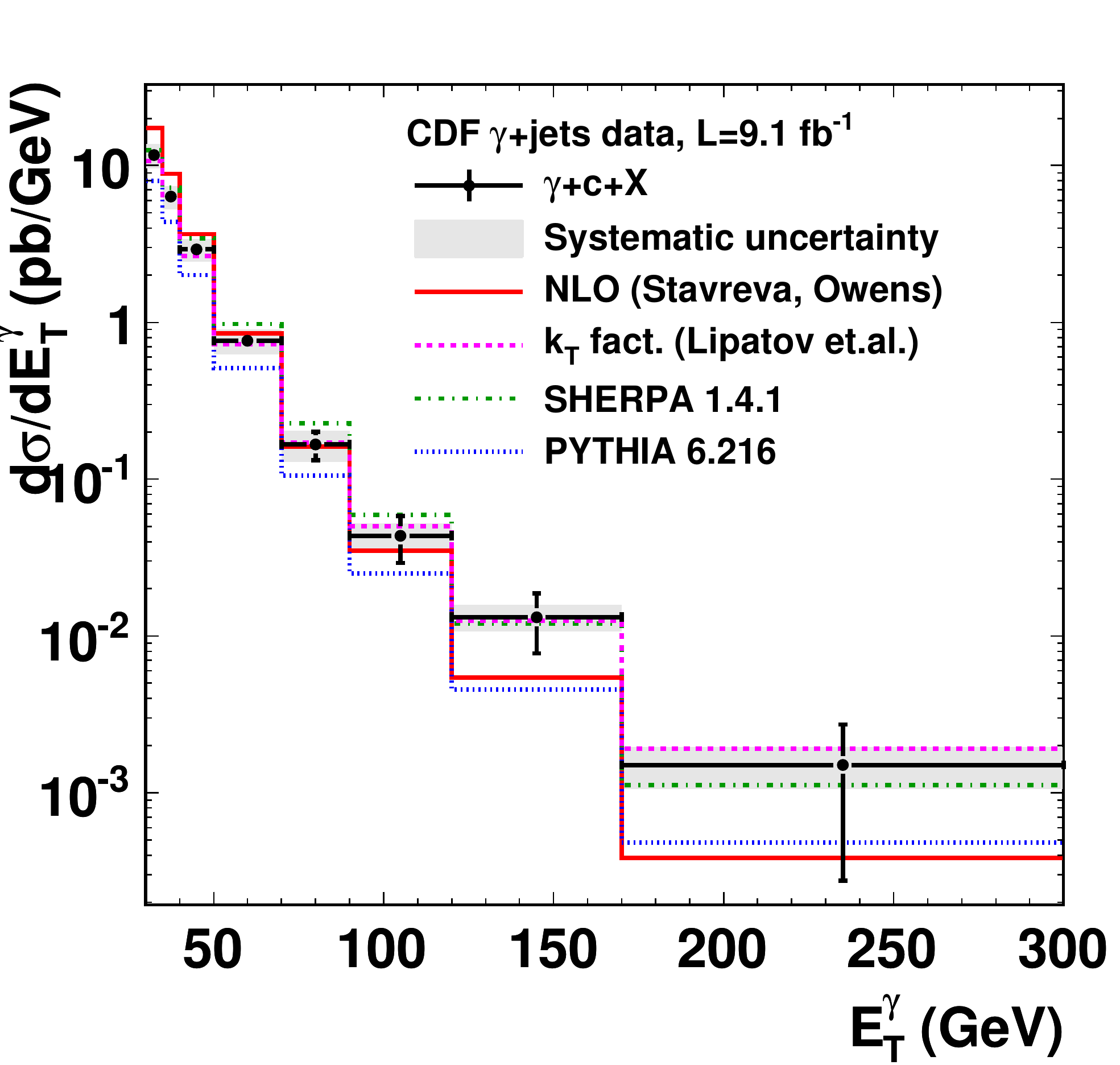}\hspace*{0cm}
\includegraphics[width=7.5cm]{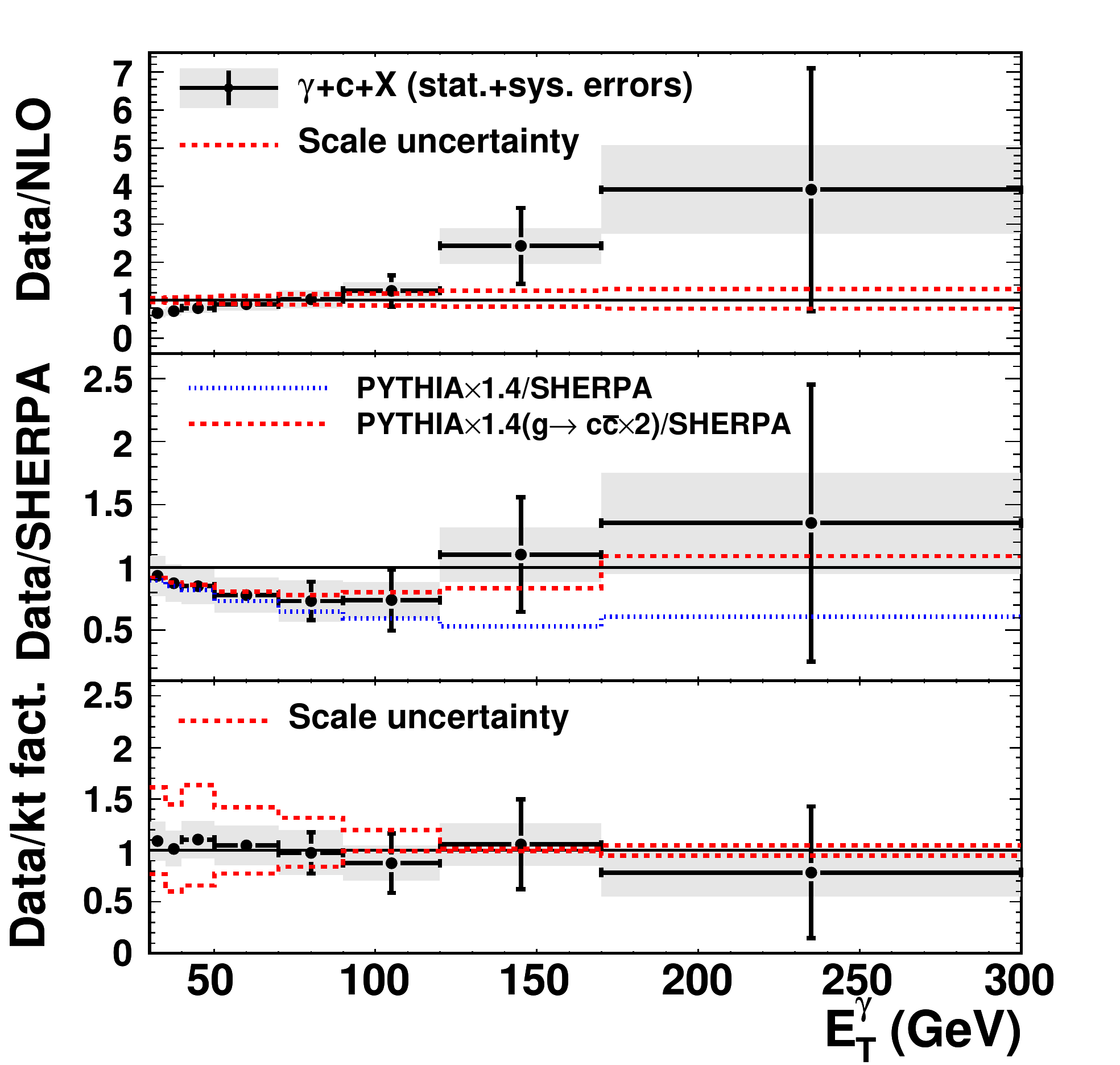}
\vspace*{-0.1cm}
\caption{The measured differential cross sections compared with theoretical predictions. The left panels show the
absolute comparisons and the right panels show the ratios
of the data over the theoretical predictions. The {\sc pythia} predictions are scaled by 1.4 in the ratio distributions. 
The comparisons are shown for $\gamma+b+X$ (top) and $\gamma+c+X$ (bottom) processes. The shaded area
around the data points indicates the total systematic uncertainty of the
measurement. The scale uncertainties are shown for the NLO and the $k_{T}$-factorization predictions.\label{fig-4}}
\end{figure*}

Figure~\ref{fig-4} shows the comparisons between the observed and predicted distributions in photon $E_{T}$. The NLO pQCD predictions agree with data at low $E_{T}^{\gamma}$ but fail to describe data for $E_{T}^{\gamma} > 70$ GeV for the bottom-jet cross section. The same trend is observed in the charm-jet cross section even though the experimental uncertainty is larger. For large $E_{T}^{\gamma}$, the dominant production process yielding a photon and a heavy quark involves a final-state gluon splitting into a heavy-flavor pair. This process is present only at leading order in the NLO calculation. The {\sc sherpa} prediction allows up to three partons in the final state, through the inclusion of additional tree-level amplitudes. The additional amplitudes also serve as a source of heavy-flavor pairs (through gluon splitting), which is important for the high $E_{T}^{\gamma}$ range. The $k_{T}$-factorization and {\sc sherpa} predictions are in reasonable agreement with the measured cross sections. The {\sc pythia} predictions disagree with the data both in rate and in shape. Scaling the {\sc pythia} prediction and doubling the rate for $g\rightarrow b\bar{b}$ or $g\rightarrow c\bar{c}$ leads to an improved agreement with the data.

\section{Conclusions}
In conclusion, we measure the differential cross sections for diphoton production and for inclusive production of a photon in association with a heavy flavor quark. The observed cross sections are compared with a wide variety of theoretical predictions. The results are important for high precision measurements of the recently discovered Higgs bosonlike particle and searches for new phenomena in diphoton final states and in final states involving the production of photons in association with heavy-flavor quarks.


\begin{thebibliography}{}
%
%
\bibitem{Aaltonen:2012jd} 
  T.~Aaltonen {\it et al.}  (CDF Collaboration),
  Phys.\ Rev.\ Lett.\  \textbf{110}, 101801 (2013).

\bibitem{Aaltonen:2013coa} 
  T.~Aaltonen {\it et al.}  (CDF Collaboration),
Phys.\ Rev.\ Lett.\  {\bf 111}, 042003 (2013).

\bibitem{Acosta:2004hw} 
  D.~Acosta {\it et al.}  (CDF Collaboration),
  Phys.\ Rev.\ D {\bf 71}, 052003 (2005).

\bibitem{Aaltonen:2011vk} 
  T.~Aaltonen {\it et al.}  (CDF Collaboration),
  Phys.\ Rev.\ D {\bf 84}, 052006 (2011).

\bibitem{Abbott:1999se}
  B.~Abbott {\it et al.} (D0 Collaboration),
  Phys.\ Lett.\  B {\bf 487}, 264 (2000);
 D.~Acosta {\it et al.}  (CDF Collaboration),
  Phys.\ Rev.\  D {\bf 71}, 092001 (2005);
  T.~Aaltonen {\it et al.}  (CDF Collaboration),
  Phys.\ Rev.\  D {\bf 78}, 072005 (2008).

\end{thebibliography}
\end{document}